# Collective Description in FELWI

D.N. Klochkov [1], K.S. Badikyan [2*]

[1] Prokhorov General Physics Institute, Moscow, Russia

[2] National University of Architecture and Construction, Yerevan, Armenia

[*] badikyan.kar@gmail.com

**Abstract:** The description of FELWI in the collective approach, is given The dispersion equations are obtained and discussed for both Raman and Thompson regimes.

## 1. INTRODUCTION

The idea to create the free electron lasers without an inversion (FELWI) was first proposed in [1], and then developed and improved in [2–5]. Specific realizations of the FELWI were proposed and considered in [6, 7]. One of the main points of the FELWI realization schemes is a proposal to use the non-collinear propagations of the electron beam and the amplified radiation. In the conventional free electron lasers (FEL) and the strophotrons these schemes are well known and have been discussed for a long time [8–18]. As applied to the FELWI with two undulators, the main idea is that at the non-collinear interaction of laser and electron beams after first undulator, there is a scattering of electrons over transversal velocities and hence of angles, and this scattering is directly related with the increase in the electron energy. Therefore, the selection of electrons over directions in the undulator gap is equivalent to the selection over the energies. In principle, it allows one in a controlled manner to change the length of the path of electrons with different energies in the inter-undulators space and the distribution of energy at the entry into the second undulator. If the devices in the inter-undulator gap have a negative dispersion property (i.e., the faster electrons spend more time on the passage of the inter-undulator space than the slow electrons), then the integral gain coefficient (over the energies of electrons) $G(\omega)$ can be positive in almost the entire region of the frequency change of the amplified wave in the vicinity of the resonance frequency of the undulator.

In the present article the analysis is performed within the framework of the multiparticle description covering both the Compton and the Raman amplification modes in the FEL.

## 2. MULTIPARTICLE DESCRIPTION

In [2–5], the model of unlimited electron and laser beams was investigated, while the real beams are limited in the longitudinal direction. Allowance for the finite widths of the latter is fundamental for lasers in which the laser and electron beams are collinear, which results in the finiteness of area of their interaction. Therefore, to take into account the finite sizes of electron and laser beams, the evaluations will be made below, in particular, the origination of the threshold in the FELWI will be analyzed. To do this, we apply the method of variance analysis to obtain the spatial amplification of the laser wave in the magnetostatic undulator with the non-collinear geometry.

## 2.1. Spatial Amplification

Consider the propagation of monoenergetic electron beam in the magnetostatic undulator. The coordinate system is chosen in such a way that the O$z$ axis coincides with the axis of a wiggler, and the vector potential of a wiggler field would be directed along the O$y$ axis. Assume, that the static magnetic field of the flat undulator $\mathbf{A}w$ is independent on the transverse coordinates $x$ and $y$ and is approximated by the harmonic function

$$\mathbf{A}_w = A_w \mathbf{e}_y = \left(A_0 e^{-i\mathbf{k}_w \mathbf{r}} + \text{c.c.}\right)\mathbf{e}_y \tag{1}$$

where $\mathbf{k}_w = (0,0,k_w)$ is the wave vector of the wiggler, c. c. denotes the complex conjugation, $\mathbf{e}_y$ is a unitary vector of $y$-axis. We assume that the linearly polarized wave $\mathbf{A}_L = A_L(t,x,z)\mathbf{e}_y = \mathbf{a}_+ e^{i(\mathbf{k}-\mathbf{k}_w)\mathbf{r}-i\omega t}$ propagates in the wiggler, where the vector potential is directed along the $y$-axis, and the wave vector is in the plane $xz$: $\mathbf{k} = (k\sin\theta, 0, k\cos\theta)$. We have neglected the Stokes wave $\mathbf{A}_- = \mathbf{a}_- e^{i(\mathbf{k}+\mathbf{k}_w)\mathbf{r}-i\omega t}$, which does not play a significant role at the resonance and should be taken into account only at the tuning out [19].

The classical dynamics of an electron in the total field of wiggler and laser wave $A = A_w + A_L A$ is described by the following Hamiltonian:

$$H = \sqrt{m^2 c^4 + c^2\left(\mathbf{P} - \frac{e}{c}\mathbf{A}\right)^2} + e\varphi = mc^2\gamma + e\varphi. \tag{2}$$

Starting from this expression, the dispersion equation was obtained in [20] for the electron beam having at the wiggler input the uniform density $nb$ and the velocity $u = (-u\sin\alpha, 0, u\cos\alpha)$

$$D_b(\omega^2 - \omega_+^2) = K^2 \omega_b^2 \gamma_0^{-3}\left(c^2 k^2 - \omega^2 + \omega_b^2 \gamma_0^{-1}\right). \tag{3}$$

Here, the following notation is used for the frequency

$$\omega_+^2 = (\mathbf{k} - \mathbf{k}_w)^2 c^2 + \frac{\omega_b^2}{\gamma_0}. \tag{4}$$

and the dispersion function of the electron beam

$$D_b = (\omega - \mathbf{k}\mathbf{u})^2 - \Omega_b^2, \tag{5}$$

which is related to the frequency of the beam $\Omega_b$, where

$$\Omega_b^2 = \omega_b^2\left[1 - (\mathbf{k}\mathbf{u})^2 / (kc)^2\right]/\gamma_0 \tag{6}$$

Here, $\omega_b^2 = 4\pi e^2 n_b / m$ is the square of the Langmuir frequency of the beam of electrons and

$$K = \frac{e}{mc^2}|A_0| \tag{7}$$

is the dimensionless amplitude of the wiggler field (the undulator parameter). The total relativistic factor of electrons is defined as $\gamma_0 = \sqrt{1+2K^2}\left(1-u^2/c^2\right)^{-1/2}$.

The dispersion equation (3) describes 4 oscillation branches $k_\upsilon = k_\upsilon(\omega)$, namely: two beam and two laser oscillation branches. At $\omega_b = 0$ these solutions have the form $(\omega - \mathbf{k}\mathbf{u})^2 = 0$ for beam waves and $\omega^2 = (\mathbf{k}-\mathbf{k}_w)^2 c^2$ for laser waves. Below we consider the solution of the dispersion equation (3) when the resonance conditions are

$$\omega = \omega_+ = (\mathbf{k}_0 \mathbf{u}) - \Omega_b, \tag{8}$$

which correspond to the maximum increment. In this case, the laser counter-propagating wave with $\omega = -|\mathbf{k}-\mathbf{k}_w|c$ can be neglected. We seek a solution in the form $\mathbf{k} = \mathbf{k}_0 + \delta\mathbf{k}$, where $\delta\mathbf{k}$ is a small complex correction to the wave vector $\delta\mathbf{k} = \mathbf{k}' + i\mathbf{k}''$. The imaginary part $\mathbf{k}''$ determines the spatial amplification of the laser wave in the undulator.

For collective regime, when $|(\mathbf{u}\delta\mathbf{k})| \ll \Omega_b$, the dispersion equation reduces to the square equation

$$\delta k^2 + \frac{K^2}{4}\frac{k_0 \Omega_b}{\gamma_0^2 u}\frac{\omega^2}{(\mathbf{k}_0\mathbf{u})^2}\left(1+\frac{\omega_b^2}{\omega\,\Omega_b\gamma_0}\right)^2 F_R(\varphi) = 0. \tag{9}$$

Where

$$F_R(\varphi) = \frac{1}{\left[\cos(\varphi-\theta) - \dfrac{k_w}{k_0}\cos\varphi\right]\cos(\varphi+\alpha)}, \tag{10}$$

$\varphi$ is an angle between the wiggler axis and the vector $\delta\mathbf{k}$. Note that the equation is valid as long as $F_R(\varphi)$ is small. Further, we will be interested in the amplification of wave along the direction of its propagation, i. e., for the case $\varphi = \theta$. In real situations $k_w \ll k_0$, so the term with $k_w/k_0$ in (10) can be neglected, then the spatial increment will be equal to

$$k'' = \frac{K}{2}\frac{k_0}{\gamma_0}\frac{\omega\sqrt{\Omega_b}}{(\mathbf{k}_0\mathbf{u})^{3/2}}\left(1+\frac{\omega_b^2}{\omega\,\Omega_b\gamma_0}\right). \tag{11}$$

Under conditions $\omega_b^2/(\omega\,\Omega_b\gamma_0) \ll 1$, when the current density of beam is small, the increment (9) has the normal dependence for the Raman regime [28]: it depends on the Langmuir frequency by law $\omega_b^{1/2}$. In the case of high-current beams, when $\omega_b^2/(\omega\,\Omega_b\gamma_0) \gg 1$, the increment has the anomalous behavior $\omega_b^{3/2}$.

For single-particle amplification (the Thomson regime $|(\mathbf{u}\delta\mathbf{k})| \gg \Omega_b$ the dispersion equation (3) is cubic:

$$\delta k^3 + \frac{K^2}{2}\frac{k_0 \Omega_b^2}{\gamma_0^2 u}\frac{\omega^2}{(\mathbf{k}_0\mathbf{u})^2}\left(1+\frac{\omega_b^2}{\omega\,\Omega_b\gamma_0}\right)^2 F_T(\varphi) = 0. \tag{12}$$

Here, $F_R(\varphi) = F_R(\varphi)/\cos(\varphi+\alpha)$. A solution of equation (12) for the imaginary part $\delta\mathbf{k}$ in the case $\varphi = \theta$

$$k'' = \frac{\sqrt{3}}{2}\left(\frac{K^2}{2}\right)^{1/3} k_0 \left[\frac{\omega\,\Omega_b}{\gamma_0(\mathbf{k}_0\mathbf{u})^2}\left(1+\frac{\omega_b^2}{\omega\,\Omega_b\gamma_0}\right)\right]^{2/3}. \qquad (13)$$

As for the Raman regime, the normal regime with $k'' \sim \omega_b^{2/3}$ is realized in the Thomson regime for lowcurrent beams $\omega_b^2/(\omega\,\Omega_b\gamma_0) \ll 1$ and the anomalous regime with $k'' \sim \omega_b^{4/3}$ for the high-current beams $\omega_b^2/(\omega\,\Omega_b\gamma_0) \gg 1$.

## 3. CONCLUSION

The multiparticle approach theory of FELWI is developed. Both the Compton and the Raman amplification modes in the FELWI are presented. It is shown, that for collective regime, the dispersion equation reduces to the square equation, and for single-particle amplification (the Thomson regime the dispersion equation is cubic.

## Acknowledgments

Authors thank KB Oganesyan for helpful discussions.